\documentclass[english, a4paper,12pt]{article}
\usepackage{amsmath,amssymb,amsfonts,latexsym,geometry,babel,graphicx,multicol,color,natbib,url,hyperref}
\usepackage[utf8]{inputenc}
\usepackage{tikz}
\usetikzlibrary{shapes,arrows,automata,mindmap,snakes,shadows,chains,matrix,positioning,scopes}

\graphicspath{{figures/}}
 
\title{Revealing the hidden structure of dynamic ecological networks}
\author{Vincent            Miele$^1$\footnote{Corresponding            author:
  \texttt{vincent.miele@univ-lyon1.fr} Author  contribution statement:
  VM and CM  developed the method and wrote the  article.  VM analysed
  the datasets.} and Catherine Matias$^2$}

\begin{document}
\DeclareGraphicsExtensions{.pdf, .jpg, .jpeg, .png}
\thispagestyle{empty}

\maketitle

\begin{center}
  $^1$ Universit\'e de Lyon, F-69000 Lyon; Universit\'e Lyon 1; CNRS, UMR5558, 
Laboratoire de Biom\'etrie et Biologie \'Evolutive,
F-69622 Villeurbanne, France.\\
$^2$ Laboratoire de  Probabilit\'es   et    Mod\`eles   Al\'eatoires,
UMR~CNRS~7599, 
Universit\'e Pierre  et Marie Curie, Universit\'e Paris
Diderot, Paris, France.
\end{center}

{\bf Abstract}
\begin{enumerate}
\item Recent  technological  advances and  long-term  data studies  provide
interaction data that can be modelled through dynamic networks, i.e 
a sequence of different snapshots of an evolving ecological network. Most often time
is  the parameter  along  which  these networks  evolve  but any  other
one-dimensional gradient (temperature, altitude, depth, humidity, $\dots$)
could be considered. 
\item Here we propose a statistical tool to analyse the underlying structure
of these networks and follow its evolution dynamics (either in time or any
other one-dimensional factor). It consists in extracting the main features of these
networks and summarise them into a high-level view.
\item We analyse a dynamic animal contact network and a seasonal food web and in both cases we show that our approach allows for the identification of a backbone organisation as well as interesting temporal variations at the individual level.
\item Our method, implemented into the \texttt{R} package \texttt{dynsbm}, can handle the
largest ecological datasets and is a versatile and promising tool for ecologists that study dynamic interactions.
\end{enumerate}

\vspace{0.5cm}
{\bf Key-words:  Dynamic networks; Network clustering; Stochastic block model; Animal contact network; Trophic network
}

\section*{Introduction}
Networks are  widely used in ecology  as they provide a  powerful tool
for    modelling    the    complex   interplay    between    ecological
entities. Depending on the context, those
entities can be different species or different individuals while
their interplay may be as diverse as trophic, competitive, cooperative
relations    or    even    contacts    measured    through    physical
proximity.  Studying these networks can help answering important ecological
questions about for e.g.  the structure of these interactions and their
robustness to external factors.

As~\citeauthor{Newman_Leicht_07} pointed out some 10 years ago \emph{``much of the current research on networks [...] aimed at answering the question: how can we tell what a network
	looks like, when we can't actually look at it?}''.
One first  answer has been  to develop and use  descriptive statistics
and  network  measures   such  as  connectance or   centrality~\citep[see][for  a  comprehensive
list]{Rayfield_11}.  This approach  considers any ecological  network as a  whole, assuming
that the  network is  homogeneous. A  next proposal  has been  to go
beyond  descriptive   statistics  and  consider   network  clustering,
i.e.  grouping entities  according  to their  common properties.  This
technique allows answering fundamental questions about any underlying 
network  structure: is  there  any  peculiar  non-random mixing  of
entities  that  would   be  a  sign  for   a  structural  organisation
\citep{kef16}?         is         there,         for         instance,
compartmentalisation~\citep{Montoya2015},    hierarchical
organisation~\citep{cla08} or nestedness~\citep{bas03}?

Nowadays, recent technological advances (sensors, GPS technology, 
$\dots$) and long-term data studies have given rise to an avalanche of
temporal data that need to be appropriately modelled.  
Data acquired over time can be aggregated within relevant time 
intervals  (days, seasons,  years, $\dots$)  and consequently  produce
snapshots of a same ecological network at different time steps.  With these new data, one can 
potentially address new ecological
questions which might not be tackled through the analysis of the static
network where  data is aggregated  over the full recording  time. In
the  same   way,  snapshots  of   an  ecological  network   along  any
one-dimensional factor (such as temperature, altitude, depth, humidity, $\dots$) may help analyse the evolution
of the network structure along this gradient~\citep{Stegen}. 
However   addressing  those  new  questions   requires  the
development of new methodological tools.  Up to now, very few proposals have been
made to handle what we call here ``dynamic networks'', namely any 
sequence  of   snapshots  of  a   same  ecological  network   along  a
one-dimensional parameter (that we most often call time). 

The two fundamental questions we will focus on here are the following: are there any
relevant statistical patterns in the dynamic network? If so, how does this
structure vary with time (or along the sequence)? In this article, we answer these two
key  points  and  argue  that  this  is  a  first  stone  for  further
understanding and predicting processes  on dynamic ecological networks
such as event spreading (infection or extinction, for instance). \\

We thus propose a statistical modelling approach to address the lack
of tools  to analyse dynamic ecological  networks~\citep{dynsbm}.  Our
approach mainly consists in  extending one of the techniques  dedicated to find
structural  patterns  in  static   networks,  now  focusing  on  their
dynamics.  We thus first present our methodological proposal, stating the key
concepts and introducing the  vocabulary required for handling dynamic
networks. It  is important to  stress that  our model is  suitable for
integrating arrivals and departures of entities (corresponding either to 
species invasion/extinction or birth/arrival/death/departure of individuals)
through the possible presence or absence of nodes at the different time
steps. Moreover, it can also deal with quantitative edges (and is not restricted to binary interactions) which are often available in
datasets.

We first illustrate our  approach through the analysis of the dynamic contact network in a colony of ants~\citep{mer13}.
Contact  networks   represent  a   relevant  proxy  to   study  animal
sociality~\citep{wey08}. In the literature, these networks may be built from field
observations  of  association   between  animals~\citep[e.g.  giraffes
in][]{car13}, trapping data~\citep[e.g. field voles in][]{dav15} and 
more     recently     and     predominantly     from     sensors-based
measurements~\citep[e.g. song  birds in][]{far15}. These data  are now
available  for large  time  periods,  ranging from  days  to years  of
observations for instance.  It is therefore possible to investigate the
(in-)stability  of  the  social  structure~\citep{pin13} and potentially question the impact of other time-related factors (seasonal changes,  response to stresses
such as draught, arrival/departure of a peculiar individual, $\dots$).

We lastly present the study of a seasonal trophic network~\citep[or {\it food
  web};][]{woo05}.  The structure of trophic relations has been intensively
studied  in  the network  framework~\cite[see][for  a  review on  food
webs]{tho12}.  Nowadays, following the seminal work of \cite{Baird},  new  datasets  allow  for   monitoring  the variation  of  this structure  along  temporal  gradients (seasons  or
years), spatial gradients~\citep[latitudinal or
longitudinal for instance;][]{kor15} or qualitative gradients~\citep[increasing habitat modification;][]{tyl07}. 
We will restrict here to dynamic
trophic networks  corresponding to  different temporal snapshots  of a
food web. In  this context, studying such structural  variation (or on
the contrary, structural stability) can  be appropriate to analyse the
system's response   to  major   changes  (species   extinctions,  environmental
perturbations, climate  change, etc).  An underlying  issue is whether
there is resilience of this structural organisation, or rather cyclic dynamics with a return to an equilibrium state.

\section*{Materials  and methods}

\subsection*{From static to dynamic networks}
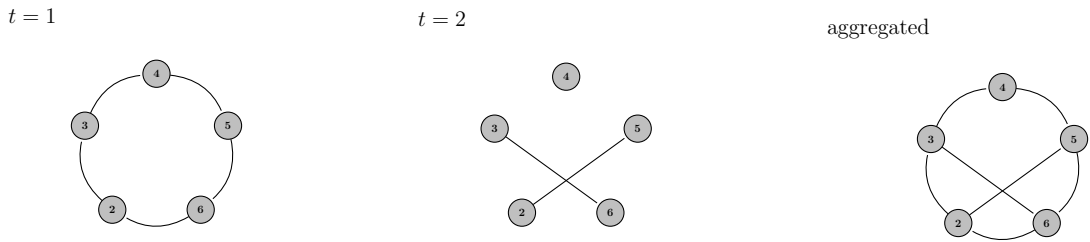
\begin{figure}
	\begin{minipage}{0.32\linewidth}
		\begin{tikzpicture}
		\tikzstyle{every node}=[fill=white]
		\tikzstyle{every state}=[text=black,scale=0.7,draw=none]
		\node[state] at (1.5,3){$t=1$};
		
		\tikzstyle{every edge}=[-,>=stealth',shorten >=1pt,auto,thin,draw]
		\tikzstyle{every state}=[text=black,scale=0.35,transform shape]

		\foreach   \name/\angle/\text   in  {B1/234/\textbf{2},   B2/162/\textbf{3},
			B3/90/\textbf{4}, B4/18/\textbf{5}, B5/-54/\textbf{6}} {
			\node[fill=lightgray,state,xshift=9cm,yshift=3.5cm]     (\name)    at
			(\angle:1cm) {\text}; 
		}
		
		\tikzstyle{every node}=[]  
		\path (B3) edge [bend left]  (B4);
		\foreach \from/\to in {1/2,2/3,4/5,5/1}{
			\path (B\from) edge [bend left] (B\to);
		}
		\end{tikzpicture}
	\end{minipage}
	\begin{minipage}{0.32\linewidth}
		\begin{tikzpicture}
		\tikzstyle{every node}=[fill=white]
		\tikzstyle{every state}=[text=black,scale=0.7,draw=none]
		\node[state] at (1.5,3){$t=2$};
		
		\tikzstyle{every edge}=[-,>=stealth',shorten >=1pt,auto,thin,draw]
		\tikzstyle{every state}=[text=black,scale=0.35,transform shape]

		\foreach   \name/\angle/\text   in  {B1/234/\textbf{2},   B2/162/\textbf{3},
			B3/90/\textbf{4}, B4/18/\textbf{5}, B5/-54/\textbf{6}} {
			\node[fill=lightgray,state,xshift=9cm,yshift=3.5cm]     (\name)    at
			(\angle:1cm) {\text}; 
		}
		
		\tikzstyle{every node}=[]  
		\path  (B2) edge (B5)
		(B1) edge (B4);
		\end{tikzpicture}
	\end{minipage}
	\begin{minipage}{0.32\linewidth}
		\begin{tikzpicture}
		\tikzstyle{every node}=[fill=white]
		\tikzstyle{every state}=[text=black,scale=0.7,draw=none]
		\node[state] at (1.5,3){aggregated};
		
		\tikzstyle{every edge}=[-,>=stealth',shorten >=1pt,auto,thin,draw]
		\tikzstyle{every state}=[text=black,scale=0.35,transform shape]

		\foreach   \name/\angle/\text   in  {B1/234/\textbf{2},   B2/162/\textbf{3},
			B3/90/\textbf{4}, B4/18/\textbf{5}, B5/-54/\textbf{6}} {
			\node[fill=lightgray,state,xshift=9cm,yshift=3.5cm]     (\name)    at
			(\angle:1cm) {\text}; 
		}
		
		\tikzstyle{every node}=[]  
		\path  (B2) edge (B5)
		(B1) edge (B4);
		\path (B3) edge [bend left]  (B4);
		\foreach \from/\to in {1/2,2/3,4/5,5/1}{
			\path (B\from) edge [bend left] (B\to);
		}
		\end{tikzpicture}
	\end{minipage}
	\caption{Same  data a)  modelled by  a two  time steps  dynamic
          network (left and centre) or b) aggregated over the whole time
          period into a static network (on the right). The structure in the static case does not reflect the complexity of the network structure which clearly varies with time. Indeed, edges present at $t=1$ and $t=2$ are disjoint.}
	\label{fig:aggregated}
\end{figure}

An  ecological network  is composed  of nodes  that correspond  to any
ecological entities -- e.g.  species,
individuals or  communities;  while
edges (or links) characterise presence/absence of an interaction between any two
entities and may be valued in some cases.
For instance, values may be the frequencies of contacts
between  two  individuals~\citep{sil11}  or the  number  of  field
observations of interactions between two species. 
When  this network  is unique and  covers an  entire time
period, it is called a {\it static} network. While many empirical
data were aggregated over a whole period of observation recording, it is
important to realise that such  aggregation could lead to an incorrect
understanding  of   the  network   structure  due  to   the  smoothing
aggregation process (cf.  Figure~\ref{fig:aggregated}). An approach to
study  the temporal  dynamics of  a set  of interactions  is the  {\it
  discrete time snapshots} approach \citep[see][for a complete perspective]{Blonder_12}.
It  consists in  aggregating  data over  specific  time frames  (days,
months, seasons, years or any  relevant frame regarding the ecological
system of  interest) and  to obtain what  \citeauthor{Blonder_12} call
{\it time-aggregated dynamic networks}. In the following, we use
the term  {\it dynamic networks} and  while we refer to  time as being
the parameter that drives the evolution, we recall that this could be
any other relevant one-dimensional factor.

Formally, we assume $T$ time steps, a number $N_t$ of nodes at each time step $t$, a total number of nodes $N$
(with $N \ll  N_1+\dots+N_t$) and edges record  the presence (possibly
valued) or absence of an interaction between any two pair
of nodes at each time step.
Note that  our set-up is  different from  the one corresponding  to the
observation of the full interaction flow, namely when data
consists   in  the   complete  knowledge   of  edges   appearance  and
disappearance along a continuous gradient. Indeed in our case data is still aggregated over some time
intervals or corresponds to a sequence of networks which are 
specific    to    a   set    of discrete   values    of   a    one-dimensional
factor. 
When considering continuous time interaction flow data, the
object of interest (the flow) is called a {\it temporal network}
\citep{Holme_review} and this setup will not be explored in this article.

Lastly, it is important to mention that the time frame selection may be 
an issue in cases where choosing the resolution for the time aggregation
is not driven  by the ecological question. Indeed, in  many cases, the
choice of  the time  frame is expert-based:  for instance  the dataset
from~\cite{rey15} consists in $T=52$ days of observation including the
breeding  season, but  it is  possible to  restrict to  $T=3$ networks
(before, during and after the breeding season) to study the network 
variations due to reproduction period. This choice might not be harmless and
for instance \cite{Blonder_12} showed that the degree distribution in networks can be sensitive to the time frame
selection~\citep[see also][for  a statistical perspective]{Rajmonda}.  
It is out of the scope of the present work to explore this frame selection problem.

\subsection*{Stochastic block models (SBM)}
\begin{figure}[h!] 
	\begin{minipage}{0.5\linewidth}
		a) \begin{tikzpicture}
		\tikzstyle{every node}=[fill=white]
		\tikzstyle{every state}=[text=black,scale=0.7,draw=none]
		
		\tikzstyle{every edge}=[-,>=stealth',shorten >=1pt,auto,thin,draw]
		\tikzstyle{every state}=[text=black,scale=0.35,transform shape]
		
		\tikzstyle{every node}=[fill=white]
		\node[fill=green,state] (A1) at (1,1) {\textbf{1}};
		
		\foreach   \name/\angle/\text   in  {B1/234/\textbf{2},   B2/162/\textbf{3},
			B3/90/\textbf{4}, B4/18/\textbf{5}, B5/-54/\textbf{6}} {
			\node[fill=yellow,state,xshift=9cm,yshift=3.5cm]     (\name)    at
			(\angle:1cm) {\text}; 
		}
		
		\tikzstyle{every node}=[]  
		\path  (B2) edge (B5)
		(B1) edge (B4);
		\path (B3) edge [bend left]  (B4);
		\foreach \from/\to in {1/2,2/3,4/5,5/1}{
			\path (B\from) edge [bend left] (B\to);
		}
		
		\tikzstyle{every node}=[fill=green]
		\node[state] (C1) at (0,0) {\textbf{7}};
		\node[state] (C2) at (1,-0.5) {\textbf{8}};
		\node[state] (C4) at (-0.5,0.5) {\textbf{10}};
		\node[state] (C5) at (-0.5,1.5) {\textbf{11}};
		\node[state] (C6) at (0,2) {\textbf{12}};
		\path (C1) edge [bend right] (C2)
		(C4) edge [bend right] (C6)
		(C4) edge (C1)
		(C4) edge [bend left] (C5)
		(C1) edge [bend left] (C6)
		(C2) edge (C6)
		(C2) edge (C5)
		(C5) edge (C6); 
		\path (C1) edge (A1)
		(C2) edge  [bend right] (A1)
		(C6) edge [bend left] (A1);
		
		\path (A1) edge [bend right] (B2);
		\end{tikzpicture}
	\end{minipage}
	\begin{minipage}{0.5\linewidth}
		b) \begin{tikzpicture}
		\tikzstyle{every node}=[fill=white]
		\tikzstyle{every state}=[text=black,scale=0.7,draw=none]
		
		\tikzstyle{every edge}=[-,>=stealth',shorten >=1pt,auto,thin,draw]
		\tikzstyle{every state}=[text=black,scale=0.35,transform shape]
		
		\tikzstyle{every node}=[fill=white]
		\node[fill=red,state] (A1) at (1,1) {\textbf{1}};
		
		\foreach   \name/\angle/\text   in  {B1/234/\textbf{2},   B2/162/\textbf{3},
			B3/90/\textbf{4}, B4/18/\textbf{5}, B5/-54/\textbf{6}} {
			\node[fill=yellow,state,xshift=9cm,yshift=3.5cm]     (\name)    at
			(\angle:1cm) {\text}; 
		}
		
		\tikzstyle{every node}=[]  
		\path  (B2) edge (B5)
		(B1) edge (B4);
		\path (B3) edge [bend left]  (B4);
		\foreach \from/\to in {1/2,2/3,4/5,5/1}{
			\path (B\from) edge [bend left] (B\to);
		}
		
		\tikzstyle{every node}=[fill=green]
		\node[state] (C1) at (0,0) {\textbf{7}};
		\node[state] (C2) at (1,-0.5) {\textbf{8}};
		\node[state] (C4) at (-0.5,0.5) {\textbf{10}};
		\node[state] (C5) at (-0.5,1.5) {\textbf{11}};
		\node[state] (C6) at (0,2) {\textbf{12}};
		\path (C1) edge (A1)
		(C2) edge (A1)
		(C4) edge (A1) 
		(C5) edge (A1)
		(C6) edge [bend left] (A1);
		
		\path (A1) edge [bend right] (B2) 
		(A1) edge [bend left]   (B3) 
		(A1) edge [bend right] (B1);
		
		\end{tikzpicture}
	\end{minipage}
	\caption{a) Network  with a  clear modular structure  with two
          modules  (green   and  yellow)  b) Network with a complex structure including different patterns: a module (yellow), a hub (red) and a set of peripheral nodes (green).  Both networks can be modelled by a SBM with different parameters that capture the structural organisation, with two and three groups respectively. Besides, it is not clear what would be the results of a modular detection algorithm on the second network (as it is not modular).}
	\label{fig:modular}
\end{figure}
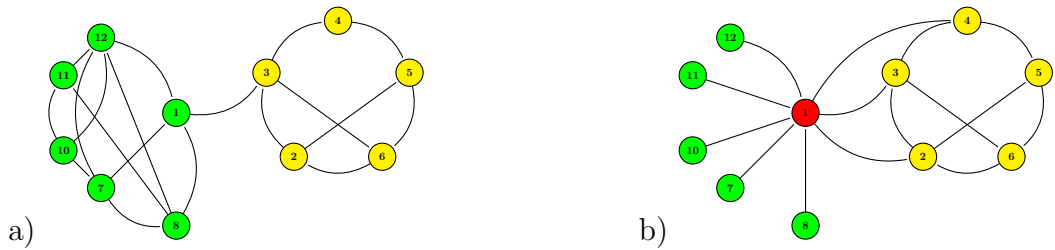

In the  field of network analysis,  one of the most  exciting research
problems  of  the   last  decade  has  been   the  network  clustering
question. Moving beyond descriptive statistics, the goal here is to propose algorithms to extract a high-level view of complex networks, i.e. zooming out the network. 
Network clustering  consists in grouping  nodes based on  their common
characteristics. It  often rhymes with  finding {\it modules}  or {\it
  communities} \citep[or {\it compartments}; see][]{sto11}. A module is a set of nodes with much more edges between these nodes than with the others. 
An important  drawback of module-based  approaches appears when,
quoting~\citeauthor{Newman_Leicht_07}, we ask: ``\emph{could there be interesting and
relevant structural features  of networks that we have  failed to find
simply because  we haven't thought  to measure the right  thing?}''. In
other words, is it relevant to search for modular structure in a
network that can be  structured in any other ways? Following this objection, methods based on
statistical inference arose which rely on the principle of grouping nodes
that have similar interaction patterns (e.g. hubs, modules, peripheral 
nodes; see Figure~\ref{fig:modular}) without any {\it a priori} knowledge. This is the purpose of a general class of models called {\it stochastic block models} (SBM).

SBM   have    been   developed   for   analysing    complex   networks
\citep{NS01,Newman_Leicht_07,Daudin_etal08,gui09,Matias_Robin_review}
and  more  recently  used  to decipher  the  structure of
ecological networks such as hosts-parasites~\citep{Mariadassou_etal},
food  webs~\citep{Picard_etal09,bas11} and multi-interactions network~\citep{kef16}.

Let $\{Y_{ij}\}_{1\leq  i,j\leq N}$  be the set  of edges  between any
possible couple of nodes $(i,j)$. The model may be defined either for directed or
undirected networks and may allow the presence of self-interactions (edges $Y_{ii}$). 
The principle of SBM is the following: we assume that the ecological entities (nodes) can be gathered into $Q$ groups based on their common interaction properties. Therefore the distribution of $Y_{ij}$ is specified conditionally on the group memberships such that
$$
Y_{ij} \sim f(\Theta_{ql}) \ | \ i \in \textrm{group } q, \, j \in \textrm{group } l
$$
where $f$  is any  probability distribution parametrised  by $\Theta$
(called interaction parameter).  The group memberships are unknown, as
well   as   the   interaction   parameters.   An   EM-like   algorithm
\citep[Expectation-Maximisation,   see][]{DLR}   allows  for   jointly
estimating   memberships  and   parameters~\citep{Daudin_etal08}.  The
statistical  procedure  finally  displays  a high-level  view  of  the
network: what  kind of interaction  patterns are present  (through the
interaction parameters  $\Theta_{ql}$) and which nodes  participate in
those patterns (through the group memberships).

A key advantage  of  SBM is  the possibility  to  plug any  probability
distribution $f$ in order to fit any kind of interactions.  For
instance, one can use a Bernoulli distribution for binary interactions~\citep{NS01}, 
a Gaussian for frequencies or a Poisson for a number of interactions~\citep[weighted
interactions;][]{Mariadassou_etal},  a multinomial  for finitely  many
values or even a multivariate distribution for
multivariate edges. One can also use the combination of any of those
distributions with  a Dirac  mass at  0 so as  to obtain  a 0-inflated
distribution  that accounts  for sparsity  in the  network~\citep[in the
valued case, not all interactions necessarily exist;][]{Ambroise_Matias}. 
Relying on a probabilistic framework allows for modelling some randomness and
variability in the observations and consequently provides robustness to possible errors or missing data.

\subsection*{Dynamic stochastic block models (dynSBM)}

How can we analyse dynamic ecological networks to extract structural information?
At the time  of writing, only a few alternatives  based on descriptive
statistics \citep{Holme_review} or on  evolving modules \citep{muc10} have
been considered.  Following the above mentioned objections, we claim 
that  a  model-based clustering  approach  could  be relevant  and  we
recently  proposed to  extend  the SBM  approach  to dynamic  networks
introducing {\it dynamic stochastic block models}~\citep[dynSBM,][]{dynsbm}.

To develop such an extension, an important question to ask is what could be the meaning of zooming out an object that can change with time?
Our answer is to capture the dynamics of a high-level view of the network.
This means tracking the evolution of the group behaviours (i.e. the
interaction parameters) as well as
the nodes group memberships with time.
Technically, we  rely on  a  collection of  SBM for
modelling the different snapshots at each time step combined with $N$ (the number
of nodes)  independent and identically distributed  Markov chains that
capture the evolution of a node group through time. 
Thus at any time step  $t$, our estimate of the group of a node $i$ depends on the SBM
estimated for the network at time $t$ and on the group of this node at
time step $t-1$. The model is now characterised through 
\begin{align*}
&Y_{ij}^t \sim f(\Theta_{ql}^t) \ | i \in \textrm{group } q , \, j \in
  \textrm{group } l \textrm{ at time } t \\
&\mathbb{P}(i \in \textrm{group } q \textrm{ at time } t \ | \ i \in \textrm{group } q' \textrm{ at time } t-1) = \Pi_{qq'} 
\end{align*}
where $\Pi$  is the  (common) transition matrix  of the  $N$ different
group memberships Markov chains.  Reconstructing the
different SBM and the common Markov chain parameters has to be done
jointly.

\begin{figure}[h!]
	\begin{minipage}{0.48\linewidth}
		\begin{tikzpicture}
		\tikzstyle{every node}=[fill=white]
		\tikzstyle{every state}=[text=black,scale=0.7,draw=none]
		\node[state] at (1.5,3){${t=1}$};
		
		\tikzstyle{every edge}=[-,>=stealth',shorten >=1pt,auto,thin,draw]
		\tikzstyle{every state}=[text=black,scale=0.35,transform shape]
		
		\tikzstyle{every node}=[fill=white]
		\node[fill=red,state] (A1) at (1,1) {\textbf{1}};
		
		\foreach   \name/\angle/\text   in  {B1/234/\textbf{2},   B2/162/\textbf{3},
			B3/90/\textbf{4}, B4/18/\textbf{5}, B5/-54/\textbf{6}} {
			\node[fill=yellow,state,xshift=9cm,yshift=3.5cm]     (\name)    at
			(\angle:1cm) {\text}; 
		}
		
		\tikzstyle{every node}=[]  
		\path  (B2) edge (B5)
		(B1) edge (B4);
		\path (B3) edge [bend left]  (B4);
		\foreach \from/\to in {1/2,2/3,4/5,5/1}{
			\path (B\from) edge [bend left] (B\to);
		}
		
		\tikzstyle{every node}=[fill=green]
		\node[state] (C1) at (0,0) {\textbf{7}};
		\node[state] (C2) at (1,-0.5) {\textbf{8}};
		\node[state] (C4) at (-0.5,0.5) {\textbf{10}};
		\node[state] (C5) at (-0.5,1.5) {\textbf{11}};
		\node[state] (C6) at (0,2) {\textbf{12}};
		\path (C1) edge (A1)
		(C2) edge (A1)
		(C4) edge (A1) 
		(C5) edge (A1)
		(C6) edge [bend left] (A1);

		\path (A1) edge [bend right] (B2) 
		(A1) edge [bend left]   (B3) 
		(A1) edge [bend right] (B1);
		
		\end{tikzpicture}
	\end{minipage}
	\begin{minipage}{0.48\linewidth}
		\begin{tikzpicture}
		\tikzstyle{every node}=[fill=white]
		\tikzstyle{every state}=[text=black,scale=0.7,draw=none]
		\node[state] at (1.5,3){$t=2$};
		
		\tikzstyle{every edge}=[-,>=stealth',shorten >=1pt,auto,thin,draw]
		\tikzstyle{every state}=[text=black,scale=0.35,transform shape]
		
		\tikzstyle{every node}=[fill=white]
		\node[fill=red,state] (A1) at (1,1) {\textbf{1}};
		
		\foreach   \name/\angle/\text   in  {B1/234/\textbf{2},   B2/162/\textbf{3},
			B3/90/\textbf{4}, B4/18/\textbf{5}, B5/-54/\textbf{6}} {
			\node[fill=yellow,state,xshift=9cm,yshift=3.5cm]     (\name)    at
			(\angle:1cm) {\text}; 
		}

		\path (A1) edge [bend left] (B3) 
		(A1) edge  (B2) 
		(A1) edge (B4) 
		(A1) edge [bend right]  (B5) 
		(A1) edge [bend right] (B1);

		\tikzstyle{every node}=[fill=green]
		\node[state] (C1) at (0,0) {\textbf{7}};
		\node[state] (C2) at (1,-0.5) {\textbf{8}};
		\node[state] (C4) at (-0.5,0.5) {\textbf{10}};
		\node[state] (C5) at (-0.5,1.5) {\textbf{11}};
		\node[state] (C6) at (0,2) {\textbf{12}};
		\path (C1) edge [bend right] (C2)
		(C4) edge [bend right] (C6)
		(C4) edge (C1)
		(C4) edge [bend left] (C5)
		(C1) edge [bend left] (C6)
		(C2) edge (C6)
		(C2) edge (C5)
		(C5) edge (C6); 
		\path (C1) edge (A1)
		(C2) edge  [bend right] (A1)
		(C6) edge [bend left] (A1);
		
		\end{tikzpicture} \\scenario 1 (not dynSBM)\\
		
		\begin{tikzpicture}
		\tikzstyle{every node}=[fill=white]
		\tikzstyle{every state}=[text=black,scale=0.7,draw=none]
		
		\tikzstyle{every edge}=[-,>=stealth',shorten >=1pt,auto,thin,draw]
		\tikzstyle{every state}=[text=black,scale=0.35,transform shape]
		
		\tikzstyle{every node}=[fill=white]
		\node[fill=red,state] (A1) at (1,1) {\textbf{1}};
		
		\foreach   \name/\angle/\text   in  {B1/234/\textbf{2},   B2/162/\textbf{3},
			B3/90/\textbf{4}, B4/18/\textbf{5}, B5/-54/\textbf{6}} {
			\node[fill=green,state,xshift=9cm,yshift=3.5cm]     (\name)    at
			(\angle:1cm) {\text}; 
		}
		
		\path (A1) edge [bend left] (B3) 
		(A1) edge  (B2) 
		(A1) edge (B4) 
		(A1) edge [bend right]  (B5) 
		(A1) edge [bend right] (B1);

		\tikzstyle{every node}=[fill=yellow]
		\node[state] (C1) at (0,0) {\textbf{7}};
		\node[state] (C2) at (1,-0.5) {\textbf{8}};
		\node[state] (C4) at (-0.5,0.5) {\textbf{10}};
		\node[state] (C5) at (-0.5,1.5) {\textbf{11}};
		\node[state] (C6) at (0,2) {\textbf{12}};
		\path (C1) edge [bend right] (C2)
		(C4) edge [bend right] (C6)
		(C4) edge (C1)
		(C4) edge [bend left] (C5)
		(C1) edge [bend left] (C6)
		(C2) edge (C6)
		(C2) edge (C5)
		(C5) edge (C6); 
		\path (C1) edge (A1)
		(C2) edge  [bend right] (A1)
		(C6) edge [bend left] (A1);
		
		\end{tikzpicture} \\scenario 2 (dynSBM)\\
	\end{minipage}
	\caption{Adding constraints  to avoid label  switching between
          time  steps.   At  time   $t=2$  there  are  two  equivalent
          alternative scenarios. In scenario 1 group 
          memberships stay constant in time and interaction 
          parameters change with time; in scenario 2 group 
          memberships  change  between  the  two time  steps  but  the
          intra-group  interactions stay  constant.  In this  latter
          case,  the   green  group  is  characterised   by  being  a
          'peripheral nodes'  group, the  yellow group is  a community
          and the red group is a hub. }
	\label{fig:switching}
\end{figure}
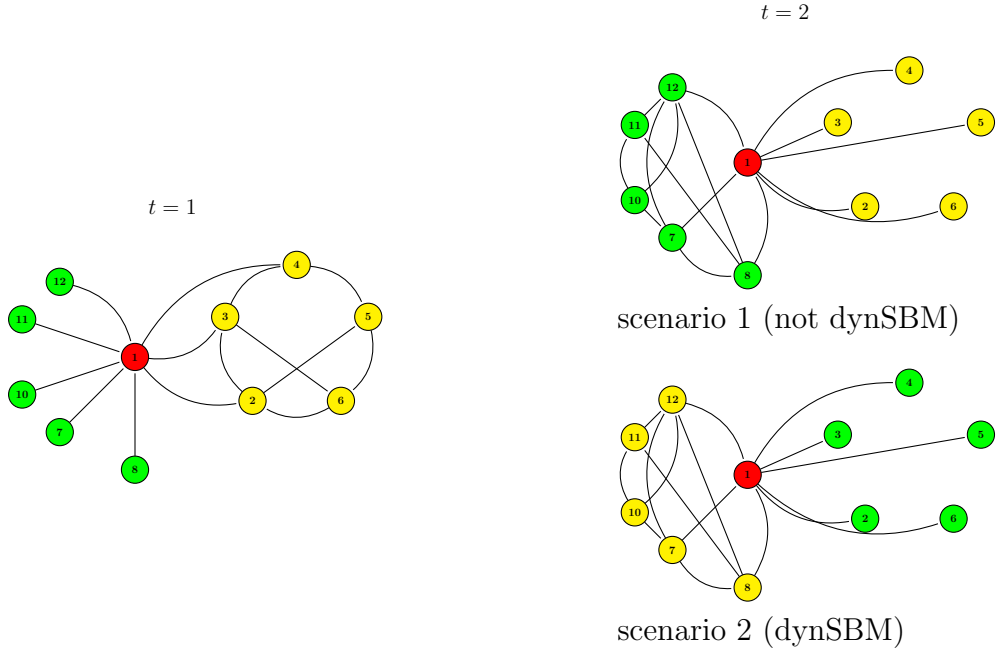

As demonstrated in \cite{dynsbm}, without adding some constraints this
model is not identifiable because of a possible label switching phenomenon
between  the  time steps  (which  is  not  the usual  label  switching
encountered in any model with latent groups). 
To illustrate this phenomenon on a toy example, let us consider a dynamic
network where the same static network is observed at two different time
steps. We assume that this network is  a {\it star} (one central node -
called the hub - is connected to all others - called peripheral
nodes).  A SBM is fitted with 2 groups. Supposing group 1 is the hub
and group 2 is composed of the peripheral 
nodes at $t=1$, two alternative scenarios are possible at $t=2$: either group 1
is still the hub and group 2 is still composed of the peripheral nodes, or the reverse
(group 1 is the peripheral nodes and group 2 the hub). Both
scenarios are  equivalent clusterings when considered  at each time
step separately (because  clusters are defined up to  a permutation of
their labels only), but globally  (meaning when considering the 2 time
steps jointly) these clusterings are different and may be fitted by dynSBM
with different parameter values that induce different interpretations.
Indeed, in  the first
scenario there are no group-switches while in the second all the nodes
undergo a group change. 
Therefore,  we need  to add  constraints to  the dynSBM  parameters to
remove this ambiguity.
In~\cite{Yang_etal_ML11}, the authors proposed to constrain the group 
memberships  to   be  constant  with   time  (see  alternative   1  in
Figure~\ref{fig:switching}).  But  this  is clearly  not  suited  to
ecological networks where  entities can evolve and move  from one role
to another; e.g. fission-fusion societies~\citep{rub15}, modification of animal
behaviour between breeding and non-breeding seasons~\citep{rey15}, response to climate
change~\citep{kor15}.  We consequently choose to allow the group 
memberships to vary freely with time but constrain some
of the interaction parameters to be stable in time (see alternative 2 in
Figure~\ref{fig:switching}). To be more specific, 
 we assume that the intra-group interactions are 
constant (namely $\Theta_{qq}^t$ does not depend on $t$, for any group
$q$). However the interaction parameters between different groups may
vary with time (namely $\Theta_{ql}^t$ depends on $t$, for any pair of
groups $(q,l)$ with $q\neq l$). This restriction is sufficient to characterise the groups and
solve the identifiability issue~\citep{dynsbm}. Note that this will also be helpful
to interpret dynSBM results as the groups are now unambiguously defined
and  meaningful: each group has its specific connectivity behaviour, with constant intra-group  and possibly
varying extra-group connectivities. 
In the meantime, the memberships are free and may change, such that any entity can change its behaviour.\\

We stress that our approach is different from a naïve one that would
separately cluster each network and use an ad-hoc procedure to resolve the label switching
problem   between  two   time  steps.    Indeed,  our   hidden  Markov
chain modelling induces dependencies  between the networks at different
time steps. As a consequence,  the clusters recovered for one specific
network use  information about  the others. We  also mention  that the
(maximal) number of groups is fixed with time though some groups might
be empty at some time steps.  This number may be selected either through a statistical
model selection criterion called ICL or relying on heuristic procedures~\citep{dynsbm}.

To summarise, the  dynSBM approach allows for  exploring the following
questions: i) Is there any high-level structure in the network, i.e. does dynSBM find
more than a single group of nodes?
ii) Does  this network structure  vary with  time, i.e. are  the nodes
group memberships evolving with time?
iii) What  are the group switches trends and frequencies,   i.e.
what are the values of the underlying Markov chain parameters? 
iv)  Are there  any stable  or  unstable individuals,  i.e. are  there
peculiar group memberships trajectories?

\subsection*{Datasets}

\paragraph{Ants interaction networks.}
Colonies  of the  ant {\it  Camponotus  fellah} were  followed with  a
tracking system that  monitored the individual positions  over days of
observations  and  dynamic  social   interactions  were  deduced  from
physical proximity~\citep{mer13}.

\paragraph{Broadstone Stream seasonal food webs.}
This dataset concerns the aquatic macro-invertebrate community of
Broadstone Stream in south-east England~\citep{food_web,woo05}. Six seasonal
connectance food  webs were  recorded, one every  two months  from May
1996  to April  1997. We  restricted here  to simple  presence/absence
information on  species (nodes) and  binary feeding links  (edges) and
did not consider quantitative data.

\section*{Results}

In a previous work, we  proposed a statistical analysis of two animals contacts
networks~\citep[sparrows        and       onagers        respectively,
see][]{dynsbm}.  Here,  we  aim  at  focusing  at  a  more  ecological
perspective and  specifically explore issues  raised by  dynSBM on
two ecological datasets: a dynamic contact network of ants~\citep{mer13} and
a seasonal food web~\citep{woo05}.

\subsection*{Dynamic animal contact networks}
The data corresponds to a colony of $N=152$ {\it Camponotus fellah} ants
observed  during $T=10$  days.  Edges  of  the resulting  dynamic
network are weighted by the  number of interactions between each pair
of ants and the network is thus undirected with no self-interactions.
After examination  of the weights  distribution, we chose to  bin those
weights into $M=3$  categories corresponding to {\it  low, medium} and
{\it high} interaction intensity. We consequently fitted a dynSBM with
a multinomial  distribution $f$ (in  fact as many multinomials  as the
number  of group  pairs $\{q,l\}$).  We  selected $Q=3$  groups with  the heuristic 
``elbow'' method \citep[see][]{dynsbm}. 

\begin{figure}[h!]
\begin{center}
	\includegraphics[width=7cm]{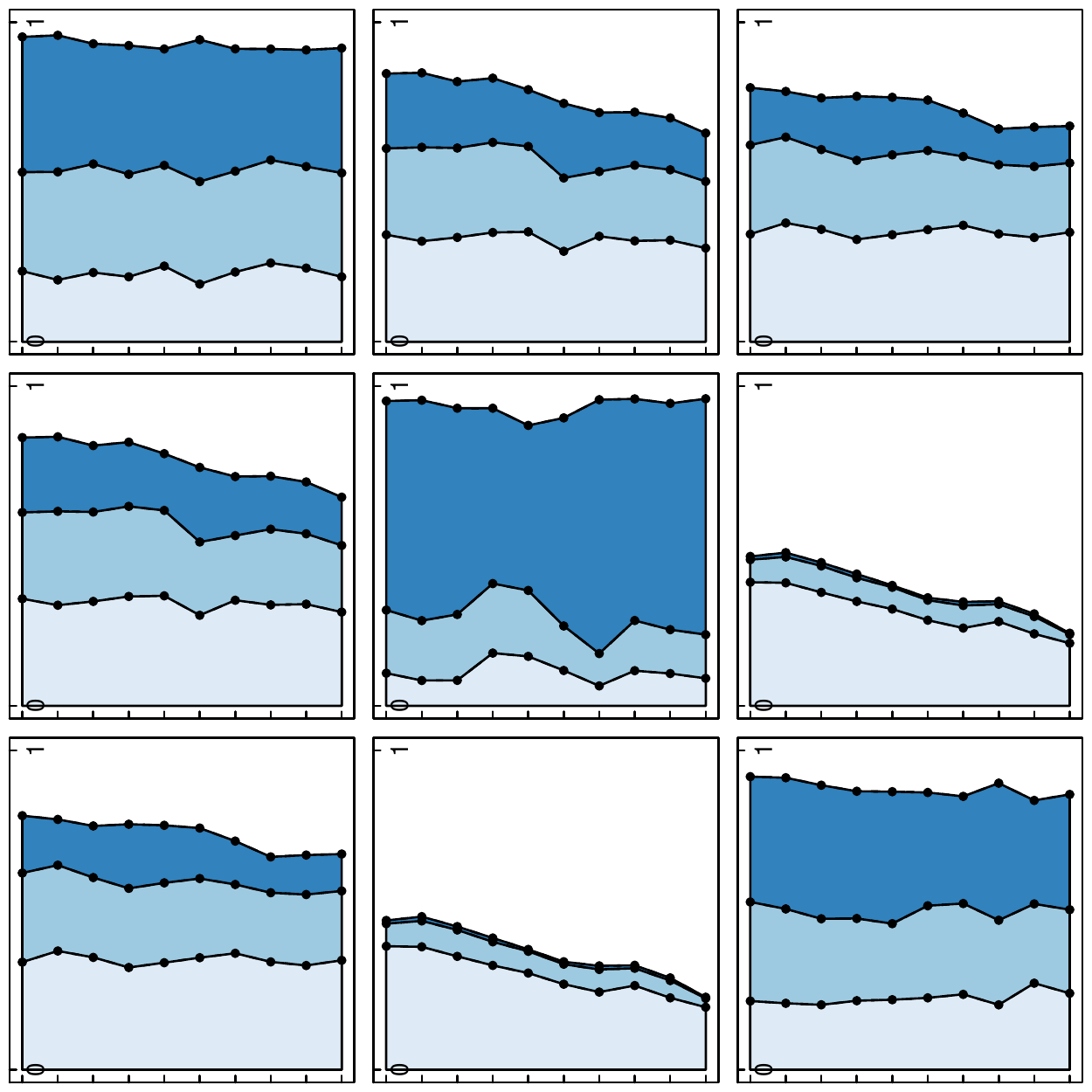}
\end{center}
\caption{Interaction properties between groups on the ants dataset.
	Interaction presence and intensity  between  nodes in  any of  the $Q=3$  groups to  the others are represented  in a global $Q\times Q$ matrix; each cell contains 
	$T=10$ time points on the $x$-axis corresponding to the different time steps. Each square represents four areas:
        the white area is the proportion of absent edges among all possible pairs of interactions; low to dark blue areas correspond to the proportion of edges (among present ones) with low to high intensity value, respectively. Plot obtained with the \texttt{connectivity.plot}  function of the \texttt{dynsbm} package.
}
\label{fig:antsconnec}
\end{figure}

We first focus on the overall structure of the dynamic network by observing the inter/intra-groups interaction
properties, as shown in the different cells of  Figure~\ref{fig:antsconnec}. Note that the global $Q\times Q$ matrix
shown here is symmetric as the network under consideration is undirected. 
 The first  key concept here is the {\it sparsity} level, {\it i.e} the amount of edges that are present over all
the possible relations (without considering edge values). We clearly see that intra-group interactions are very
frequent, in particular in groups 1 and 2 where almost any pairs of ants of these groups are in contact
(Figure~\ref{fig:antsconnec}, large blue areas in diagonal plots). This pattern is stable in time (10 days,
$x$-axis in Figure~\ref{fig:antsconnec}).
The most interesting trend about inter-groups interaction concerns group 3 which contains ants that interact with those
of group 1 but much less with those of group 2 (Figure~\ref{fig:antsconnec}, smaller blue areas in cells (2,3) or 
(3,2) than in cells (1,3) or (3,1)). These properties are the key factor determining the group boundaries ({\it i.e} the memberships) as the other inter-group interactions remain frequent. 
The next key  notion is the {\it intensity} level  that focuses on the
values of present edges and reflects the point to which ants of two given groups are more likely
to be in contact with low to high intensity. Interestingly, when two ants of group 2 are in contact (edge is present),
they are likely to be in contact with a high intensity/frequency value (Figure~\ref{fig:antsconnec}, large dark blue
area in cell (2,2)). On the contrary, even if some  contacts exist between these ants and those of group 3, which is already
unusual, these sporadic contacts exhibit low intensity (Figure~\ref{fig:antsconnec}, larger light blue area in cell (2,3)
or (3,2)). Here, the remaining intra/inter-group intensity levels do not reveal any other interesting pattern (equal proportion of intensity categories).
With all these observations, we deduce that group 2 is a so-called module (highly intra-group connected ants) relatively
disconnected from group 3 and that group 1 gathers ants ``at the interface'' {\it i.e.} interacting with partners from
any of the three groups.

\begin{figure}[h!]
	\begin{center}
	\includegraphics[width=16cm]{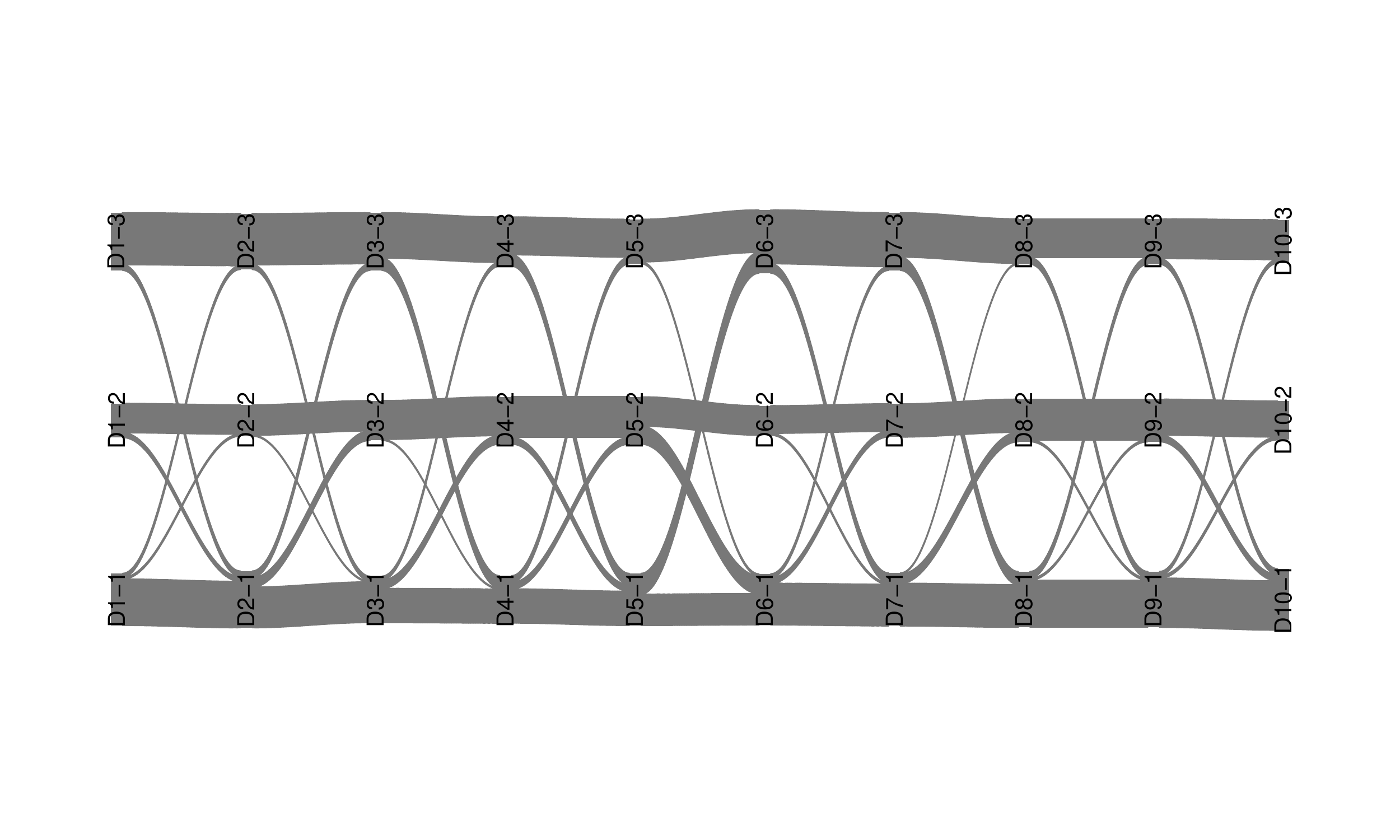}
	\end{center}
\caption{Alluvial  plot showing  the dynamics  of the  group memberships on the ants dataset.  Between two days ($t=1, \dots,10$ on the $x$-axis), each line is a flux that represents
  the switch  of one or  more ants from  a group to  another group
  ($q=1, \dots, Q$ represented on the $y$-axis).  Here, $Dt-q$ denotes group $q$ from day $t$. 
The thickness of each line is  proportional  to the  corresponding counts.
Plot obtained with the  \texttt{alluvial.plot} function of the \texttt{dynsbm} package. 
}
\label{fig:antsalluvial}
\end{figure}

We now investigate whether there are some interesting trends in the turnover of group membership. In other words, we
wonder whether all ants have the same propensity to move from one group to another one. 
We first observe that the global group turnover (i.e. the amount of group switches)
is low: 46\% of ants never switch group. 
Moreover, there are 
no group switches between groups 2 and 3 (Figure~\ref{fig:antsalluvial}, no fluxes between these groups over time). This observation, along with
the  low level of interactions  between  group 2  and  3 that we discussed before, suggests  the
existence  of  a  ``barrier''  between  these groups  that  could  be  a
consequence of  space positioning.  Indeed,~\cite{mer13}  showed that
ants were distributed over three social groups (obtained by analysing each daily static network and combining those analyses)  with different  interaction patterns  and that  there existed  some spatial segregation of the groups. We thus propose to compare our groups obtained with
dynSBM  (which  are   evolving  with  time)  and   the  social  groups
of~\cite{mer13} (which are fixed with time).  Focusing on the  ants
that  stay in the same group at  least 8 days over 10 ($111$ ants over $N=152$), we note a quasi-perfect match  between~\citeauthor{mer13} groups  and our  groups (see
Table~\ref{tab:ants}).  The modular  group 2  corresponds to  the {\it
  foragers}  of~\citeauthor{mer13}, while  the  other groups  1 and  3
correspond to the {\it cleaners} and {\it nurses} respectively. Besides retrieving this functional group, we provide another relevant information: it is now possible to study  ants playing  different social roles  over time, i.e. those that experience group switches at  certain time points.  Indeed, our dynSBM groups  allow to pinpoint these interesting  individuals that  modified their  behaviour over time and that can be of peculiar interest for specialists of the  {\it  Camponotus  fellah} system.

\begin{table}
\caption{Contingency   table   between~\citeauthor{mer13}   functional
  groups and  our dynSBM  groups (restricted to  75\% of  ants staying
   at least 8 over 10 days in the same group).}
\begin{center}
\begin{tabular}{c|ccc} 
& cleaners & foragers & nurses\\
\hline
dynsSBM group 1 & 29 & 1  & 4\\
dynsSBM group 2 & 2  & 29 & 0\\
dynsSBM group 3 & 0  & 0  & 42
\end{tabular} 
\end{center} 
\label{tab:ants}
\end{table}

\subsection*{Broadstone Stream seasonal food webs}
The number of sampled species of this aquatic  macro-invertebrate community  varies seasonally (up to $N=26$ in total
including  $10$  predators)   as  well  as  the   number  of  directed
links.  This  dataset  forms  a dynamic  trophic  network  with  $T=6$
snapshots (May, August, October,  December 1996, February, April 1997)
and we are interested in addressing  the temporal variation in the web
structure. Five species were not sampled each month but this situation
where nodes  are present/absent  over time is  supported by  our model
\citep[see supplementary material of][for details]{dynsbm}. It is also
important to mention that  self-interactions (cannibalism) exist for 6
out of 10  predator species. Again,  our model allows  for this  behaviour which
might distinguish predators among them and structure the network.  We then fitted a dynSBM with Bernoulli distributions $f$ and we selected $Q=4$ groups with the ICL criterion~\citep[see][]{dynsbm}.

\begin{figure}[h!]
	\begin{center}
		\includegraphics[width=8.5cm]{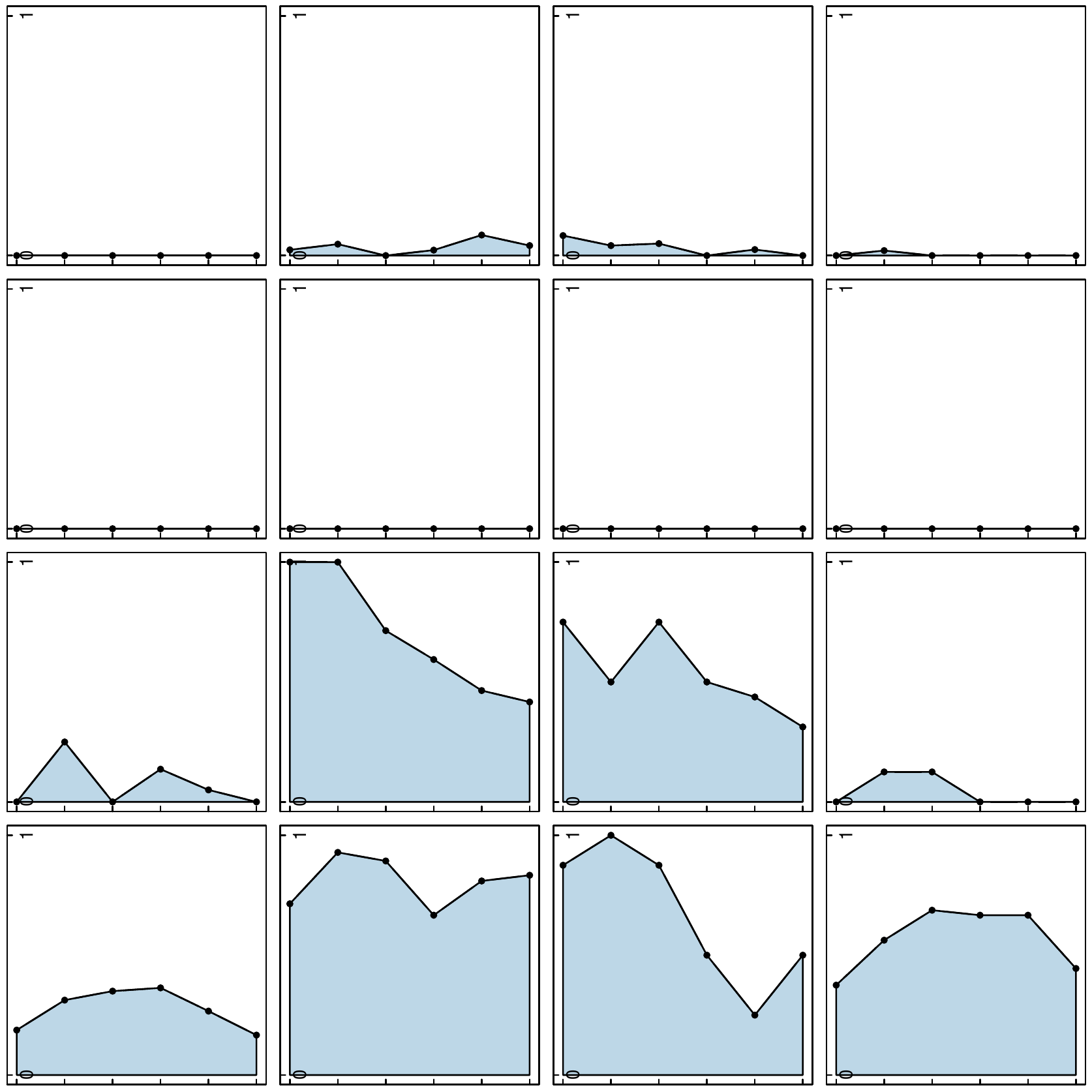}
	\end{center}
	\caption{Interaction properties between groups on the food web
          dataset.  Same  as   Figure~\ref{fig:antsconnec}  for  $Q=4$
          groups and $T=6$ time steps.  In this case, only interaction
          presence is shown (blue area) as we consider binary edges. Moreover, the $Q\times Q$ matrix shows directed interaction from group $q$ (lines) to group $l$ (columns).
	}
	\label{fig:foodwebconnec}	
\end{figure}

The    inter/intra-groups    interaction     properties    shown    in
Figure~\ref{fig:foodwebconnec}  are  not   symmetric  as  we  consider
directed networks. Therefore,  for  each pair  of  groups $\{q,l\}$  their
interaction characteristics  are twofold:  how often species  of group
$q$ eat those of group $l$, and the reverse.  As such, group 4 is
composed of  {\it omnivorous}  species that eat  species of  any other
groups, but  are only eaten  by species of their  own group (this
includes cannibalism).  Group 3 has  the same properties than  group 4
with a significant difference: species of group 3 do not eat those
of group 4.  We conclude that species from group 3 occupy intermediate positions in
 food chains whereas those of group 4 are top predators. Indeed, group
 4 is mainly composed of the  three largest species (the top predators
 {\it  Cordulegaster  boltonii},  {\it  Sialis  fuliginosa}  and  {\it
   Plectrocnemia  conspersa}) whereas  group 3  contains mostly  three
 small  species  (the larvae  of the tanypod midges {\it  Macropelopia nebulosa},
 {\it     Trissopelopia     longimana}    and     {\it     Zavrelimyia
   barbatipes}). Group 2 overall gathers preys that are mostly 
 eaten by predators of groups 3 and 4.  Species from group 1 are
 ``hidden'' species:  they do  not eat  much, and  are not  much eaten
 either. This  group is obtained  by our statistical procedure  due to
 fewer feeding links from/to other  species compared to species of the
 other groups. It is not coherent from a taxonomic point of view as it gathers a mixture of predators with little activity and secondary preys that we  call {\it peripheral species}.
Lastly, our model can deal with the overall  decrease in the number of links
after   October   (see   Figure~\ref{fig:foodwebconnec},   blue   area
decreasing with time in boxes) which is partly due to the fact that tanypods become less predatory
and more detritivorous after autumn~\citep{woo05}.

\begin{figure}[h!]
\begin{center}
	\includegraphics[width=16cm]{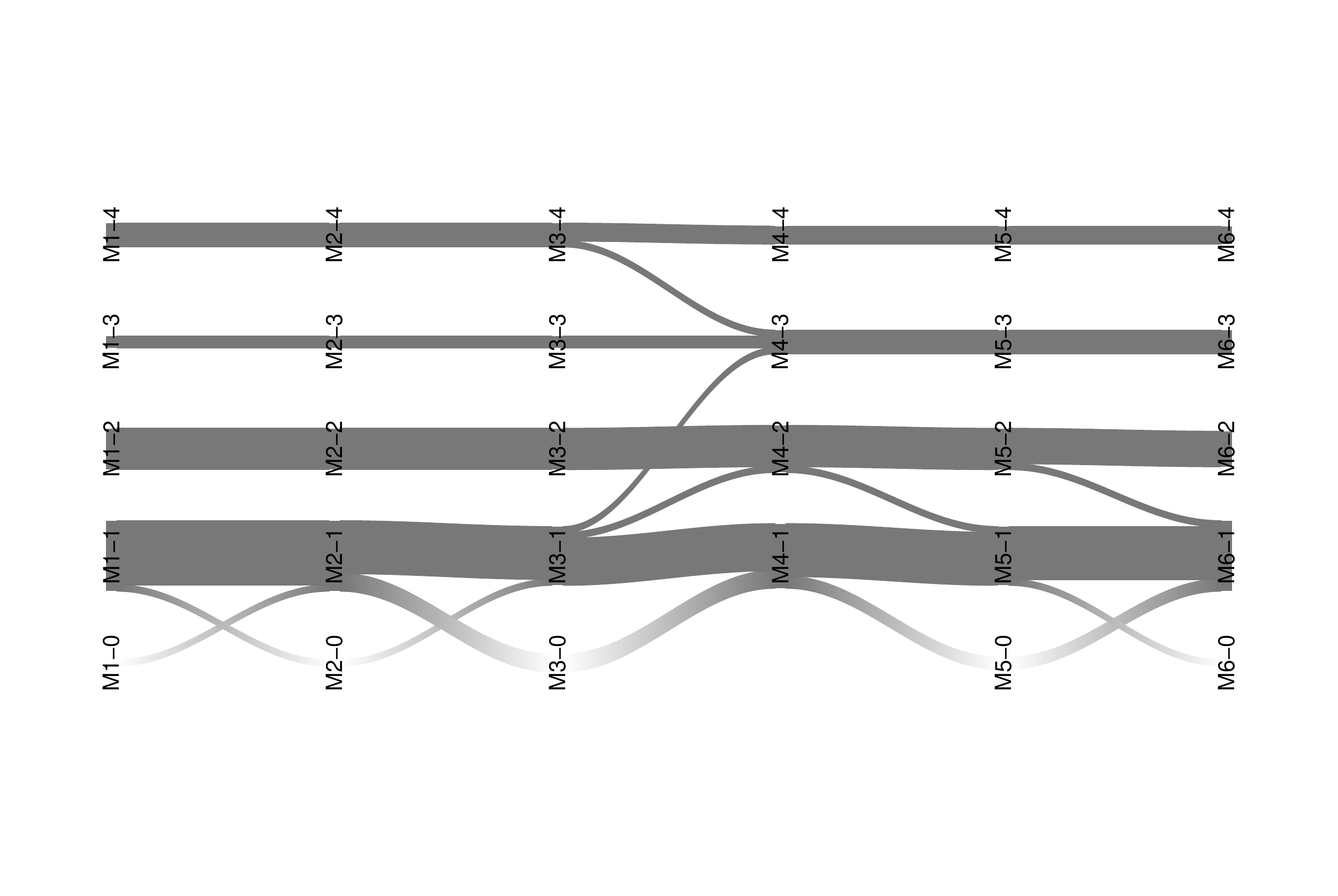}
\end{center}
\caption{Alluvial plot  showing the dynamics of  the group memberships
  on the food web  dataset.  Same as Figure~\ref{fig:antsalluvial} for
  months $t=1, \dots,6$.  Here, $Mt-q$ denotes group $q$  on month $t$
  (for $1\le q \le 4$) and special group $0$ gathers absent entities at each time step.
}
\label{fig:foodweballuvial}
\end{figure}

Now, we explore  whether species positions in the  food chains (namely
being top or intermediate predators, peripheral species or common preys) evolve or
stay constant across the seasons. We do not expect a low-level prey to
become a top predator, but the group boundaries may change due to
seasonal diet variations; for instance, {\it Macropelopia nebulosa}
eats lots  of {\it Nemurella pictetii}  in August but not  in April as
this species becomes too large; see~\cite{woo05}. 
Figure~\ref{fig:foodweballuvial} shows  that group  memberships remain
stable before winter, but some changes are observed between
October and  December. In particular,  the tanypod species  {\it Macropelopia
  nebulosa} belongs  to group 4 and  changes to
group 3 in winter.  
Indeed,  in summer  and autumn only, this species diet is similar to the one of the other members  of group  4 (the  three competitive  top predators,  that eat
each  other)  while  also  being their  prey. 
Still  between October  and December,  the stonefly  {\it Siphonoperla
  torrentium} becomes an active predator (with only 1 prey in
October and 5 in December) and moves from group 1 to group 3. The prey
{\it Prodiamesa olivacea} becomes commonly  eaten during winter and is
consequently integrated into group 2 during this period (and moves back
to group  1 in April). Lastly,  we observe that {\it  Brillia modesta}
changes from  group 2 to group  1 between December and  February: this
species becomes  the exclusive prey  of the top predators  during this
period whereas it is a common prey during the other months.

\section*{Discussion}
The inclusion of time in network  analysis has been a recent challenge
that requires ad-hoc modelling approaches. The success of these 
approaches has to be measured  by their ability to extract substantial
additional  information that  would not  be caught  by a  traditional
static network  analysis. To this aim,  we propose the use  of our new
dynamic  stochastic  block  model   to  decipher  temporal  trends  or
temporary patterns in dynamic ecological networks.

On the ants interaction network dataset, while the overall group behaviour trends are captured
by   our    model,   different   individual   behaviours    are   also
highlighted. This way, our results can be interpreted at different scales. On the food web dataset, our model underlines a clear trophic organization but also seasonal differences in the prey assemblage. These results require further investigation by experts, but it is interesting to note that our approach can play a key role in extracting unexpected patterns.

Our model is grounded  on a rigorous statistical method~\citep{dynsbm}
and   is   implemented   in  an   efficient   \texttt{R/C++}   package
that can handle hundreds to thousands of nodes. It is henceforth one of the very first tools for ecologists
facing the recent  availability of time-ordered  datasets or that
would like to explore the evolution of ecological networks with respect
to a one-dimensional factor.

\section*{Acknowledgements}
The authors would like to thank Sonia K\'efi for helping us in finding
the food web dataset and for her comments on earlier versions of this work.

\section*{Data and code accessibility}
The ants dataset is available at \url{http://datadryad.org/resource/doi:10.5061/dryad.8d8h7}.
The food web dataset is available in plain text in Table~2 of \cite{woo05}.
The R software package \texttt{dynsbm} is available at \url{http://lbbe.univ-lyon1.fr/dynsbm}.

\bibliographystyle{apalike}
\bibliography{mee_miele}

\end{document}